\documentclass[journal]{IEEEtran}
\usepackage{array}
\usepackage{textcomp}
\usepackage{stfloats}
\usepackage{url}
\usepackage{verbatim}
\usepackage{graphicx}
\usepackage{cite}
\usepackage{xcolor}
\usepackage[acronyms,nonumberlist]{glossaries}

\newacronym{BS}{BS}{base station}
\newacronym{FAS}{FAS}{fluid antenna system}
\newacronym{CFAS}{CFAS}{continuous fluid antenna system}
\newacronym{DFAS}{DFAS}{discrete fluid antenna system}
\newacronym{FA}{FA}{fluid antenna}
\newacronym{FPA}{FPA}{fixed-position antenna}
\newacronym{6G}{6G}{sixth-generation}
\newacronym{IoE}{IoE}{Internet-of-Everything}
\newacronym{MIMO}{MIMO}{multiple-input-multiple-output}
\newacronym[plural=DoF,
            longplural={degrees-of-freedom}]{DoF}{DoF}{degree-of-freedom}
\newacronym[plural=UEs,
            longplural={user equipment}]{UE}{UE}{user equipment}
\newacronym{MRC}{MRC}{maximum-ratio combining}
\newacronym{SNR}{SNR}{signal-to-noise ratio}
\newacronym{SINR}{SINR}{signal-to-interference-plus-noise ratio}
\newacronym{RF}{RF}{radio-frequency}
\newacronym{1D}{1D}{one-dimensional}
\newacronym{2D}{2D}{two-dimensional}
\newacronym{3D}{3D}{three-dimensional}
\newacronym{OP}{OP}{outage probability}
\newacronym{SISO}{SISO}{single-input-single-output}
\newacronym{LCR}{LCR}{level crossing rate}
\newacronym{AFD}{AFD}{average fade duration}
\newacronym{SIR}{SIR}{signal-to-interference ratio}
\newacronym{SER}{SER}{symbol error rate}
\newacronym{BER}{BER}{bit error rate}
\newacronym{ZF}{ZF}{zero-forcing}
\newacronym{ML}{ML}{maximum likelihood}
\newacronym{MLE}{MLE}{maximum likelihood estimation}
\newacronym{ISAC}{ISAC}{integrated sensing and communication}
\newacronym{UAV}{UAV}{unmanned aerial vehicle}
\newacronym{RIS}{RIS}{reconfigurable intelligent surface}
\newacronym{CF}{CF}{cell-free}
\newacronym{HSP}{HSP}{high {\gls{SNR}} probability}
\newacronym{EEC}{EEC}{expected Euler characteristic}
\newacronym{MMSE}{MMSE}{minimum mean-squared error}
\newacronym{CSI}{CSI}{channel state information}
\begin{document}

\title{Spatial Limits of Fluid Antenna Systems}

\author{Amy S. Inwood,~\IEEEmembership{Member,~IEEE,} Peter J. Smith,~\IEEEmembership{Fellow,~IEEE,} \\ Rajitha Senanayake,~\IEEEmembership{Member,~IEEE,}  and Michail Matthaiou,~\IEEEmembership{Fellow,~IEEE}

\thanks{This work was supported by the U.K. Engineering and Physical Sciences Research Council (EPSRC) grant (EP/X04047X/2) for TITAN Telecoms Hub. The work of P. J. Smith was supported by the Marsden Fund Council from New Zealand Government funding, managed by Royal Society Te Apārangi. The work of M. Matthaiou was supported by the European Research Council (ERC) under the European Union’s Horizon 2020 Research and Innovation Programme (grant agreement No. 101001331).}
\thanks{A. S. Inwood and M. Matthaiou are with the Centre for Wireless Innovation (CWI), Queen’s University Belfast, Belfast BT3 9DT, U.K. (e-mail: \{a.inwood, m.matthaiou\}@qub.ac.uk).}
\thanks{P. J. Smith is with the School of Mathematics and Statistics, Victoria University of Wellington, Wellington, New Zealand (e-mail: peter.smith@vuw.ac.nz).}
\thanks{R. Senanayake is with the Department of Electrical and Electronic Engineering, University of Melbourne, Melbourne, VIC 3010, Australia (email: rajitha.senanayake@unimelb.edu.au).}

}

 \maketitle

\begin{abstract}
Continuous fluid antenna systems (CFASs) represent an upper bound on the spatial diversity performance of fluid antenna systems (FASs), achieved when antennas may be positioned anywhere within a defined spatial region. This article examines the fundamental relationships governing CFAS performance. The focus is on the probability that the signal-to-noise ratio (SNR)  exceeds a prescribed high threshold, termed the high SNR probability (HSP). This is among the few FAS performance metrics that admit the derivation of closed-form expressions. Following a survey of recent analytical advances in FAS performance limits, a dimensional scaling law derived for the HSP of a single-user, single-antenna CFAS is examined. This law is then applied to the per-user high signal-to-interference-plus-noise ratio (SINR) probability of a two-antenna, two-user CFAS employing minimum mean-squared error (MMSE) combining. For both scenarios, performance gains are shown to increase consistently with both dimensionality and region size. Remarkably, the scaling law remains accurate in the two-user case, showing that, in both scenarios, the influence of additional dimensions is dominated by the CFAS size and considered threshold. Moreover, the per-user high SINR probability of the two-user system exceeds the single-user HSP, despite the addition of inter-user interference.
\end{abstract}

\begin{IEEEkeywords}
 Continuous fluid antenna systems, dimensional scaling laws, fundamental performance limits.
\end{IEEEkeywords}

\section{Introduction}
As \gls{6G} networks move from concept to design and deployment, applications and use-cases once confined to science fiction are poised to become a reality. The International Telecommunications Union's framework for IMT-2030 envisions usage scenarios spanning immersive communications, holographic communications, virtual reality, and smart cities populated by vast ecosystems of \gls{IoE} devices \cite{itu_framework_2023}. Realizing these scenarios will require substantially greater bandwidth and higher carrier frequencies than those employed in preceding generations. However, operation at higher frequencies introduces additional propagation challenges, such as higher path losses and an increased likelihood of blockages, which makes the exploitation of spatial diversity not only beneficial but also necessary.

The harnessing of spatial diversity in communications systems has a long history, tracing its origins to receive-antenna combining in the mid-twentieth century before undergoing a fundamental transformation with the formalization of \gls{MIMO} technology in the late 1990s. \gls{MIMO} technology has revolutionized communications, enabling spatial multiplexing to increase spectral efficiency, beamforming to focus radiated energy and suppress interference, and diversity combining to improve link reliability. However, the spacing between antennas has typically remained in the order of half of a wavelength. Although a reduction in this value would allow for more \glspl{DoF}, the complexity and costs of mitigating or exploiting the resulting mutual coupling between antennas have remained prohibitive.

The additional \glspl{DoF} afforded by exploiting the spatial domain at finer granularity may prove essential to realizing the ambitious performance targets of \gls{6G}. One promising technology proposed to deliver this is the \gls{FAS}. First proposed in their current form in \cite{wong_fluid_2021}, \glspl{FAS} have since emerged as an umbrella term encompassing any system in which antenna position is a reconfigurable \gls{DoF}. This could consist of \glspl{FA} with dielectric holders, mechanically movable antennas, or pixel-based switching \cite{new_tutorial_2025}. Early work focused on systems in which a single \gls{FA} could be repositioned across a discrete set of ports within a linearly constrained space \cite{wong_fluid_2021}. A key advantage of this architecture is the absence of mutual coupling constraints, permitting port spacings far below those practical in conventional arrays. This allows rich spatial diversity across a compact region, decoupling diversity gain from the hardware complexity that typically accompanies it in multi-antenna systems.

The theoretical upper bound on \gls{FAS} performance is achieved when the \gls{FA} may be positioned anywhere within a continuous space, henceforth referred to as a \gls{CFAS}. Consequently, a \gls{CFAS} of any region size\footnote{Region refers to the space within which a \gls{FA} or \glspl{FA} can move, the measure of which defines the \gls{FAS} size.} can achieve an arbitrarily small \gls{OP} and outperform \gls{MRC}, as deeper fades are spatially narrower and thus require only a small positional displacement to avoid.

This article investigates the fundamental relationships governing \gls{FAS} performance by focusing on the \gls{CFAS} as a canonical limiting case. Beyond characterizing the behavior of truly continuous systems, this perspective yields upper bounds for discrete port architectures, offering insight into how closely \gls{DFAS} designs can approach the spatial diversity limit. To begin, the fundamentals of both \gls{DFAS} and \gls{CFAS} structures is discussed. To characterize the current state of knowledge, an overview of recent research advances in the performance limits of \glspl{FAS} is presented. Attention then turns to the probability that the \gls{SNR} exceeds a prescribed high threshold, termed the \gls{HSP}. This is a key metric of \glspl{FAS} by virtue of being among the very few that admit closed-form expressions. Existing studies, such as \cite{smith_dimensional_2025}, have been restricted to single-\gls{UE} systems. While closed-form results are generally believed to be intractable for more complex systems, we investigate the extent to which single-\gls{UE} results extend to a \gls{CFAS} with two antennas and two \glspl{UE}, marking an important step toward the characterization of more complex \glspl{CFAS}. Finally, open research problems are identified and directions for future work are outlined.

\section{Background and Fundamentals}

The definition of what constitutes a \gls{FAS} is broad, and consequently there are a wide range of proposed implementation strategies and modes of operation. 

\begin{figure*}[t]
\normalsize
\includegraphics[trim={1cm 16.75cm 35cm 0cm},clip,width=\textwidth]{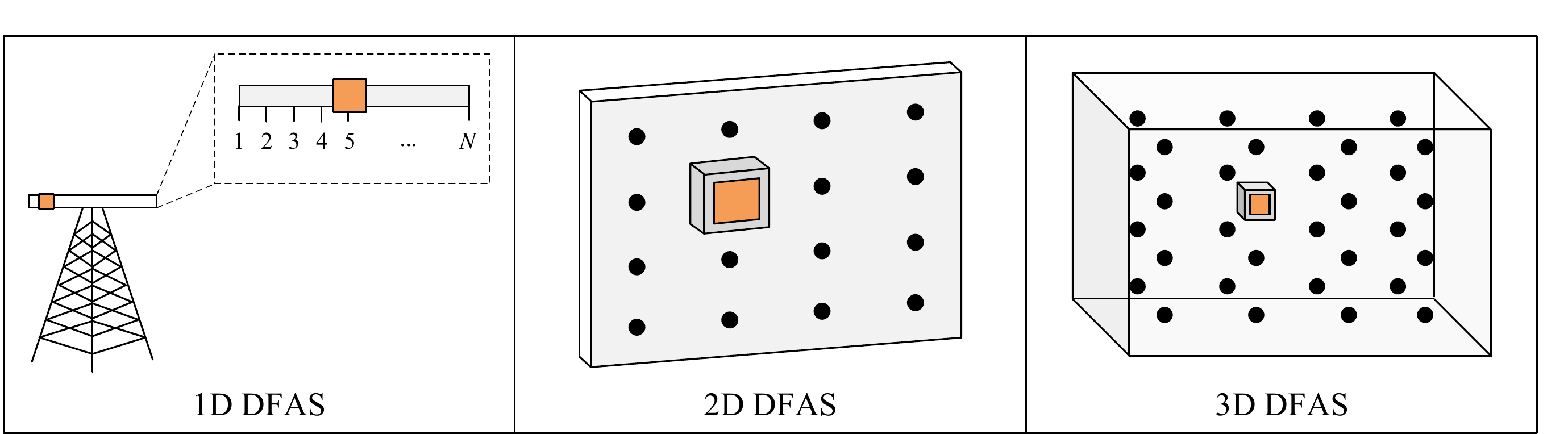}
\caption{Layout of single-antenna \glspl{DFAS} in 1-3D.}
\label{fig:DFAS}
\includegraphics[trim={1cm 16.75cm 35cm 0cm},clip,width=\textwidth]{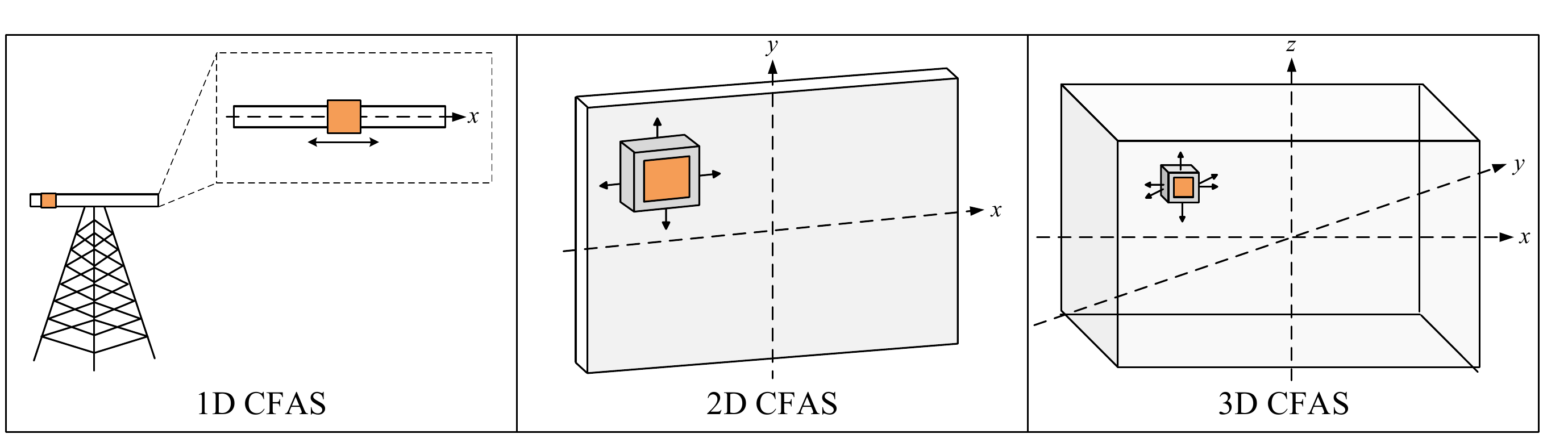}
\caption{Layout of single-antenna \glspl{CFAS} in 1-3D.}
\label{fig:CFAS}
\hrulefill
\end{figure*}

In single-antenna \glspl{DFAS}, a single \gls{FA} is connected to one of a discrete set of ports, typically equally spaced at sub-wavelength intervals. These ports may be arranged along a line, across a surface, or throughout a volume, yielding \gls{1D}, \gls{2D}, and \gls{3D} architectures respectively, as illustrated in Fig.~\ref{fig:DFAS}. The \gls{RF} chain is connected to all ports simultaneously, enabling the \gls{FA} to be served from whichever port offers the most favorable channel conditions. The \gls{CFAS} counterpart relaxes the discrete port constraint entirely, and the \gls{FA} is free to occupy any position within an equivalent spatial region. The \gls{RF} chain is connected directly to the antenna, e.g., via a flexible cable, and moves with it continuously through the spatial region. Different \gls{CFAS} architectures are shown in Fig. \ref{fig:CFAS}.

A large amount of analytical work to date has considered \gls{1D} \glspl{FAS} \cite{wong_fluid_2021,psomas_continuous_2023,new_fluid_2024,zhu_on_2025,zhu_fluid_2025,wong_performance_2020}. However, higher-dimensional \glspl{FAS} can provide significant performance improvements over their \gls{1D} counterparts. The relationships between the performance obtainable across spatial regions of varying dimensionality offer important insights and aid in system design, so an arbitrary number of dimensions is considered here.

In \glspl{DFAS}, it is highly unlikely that any port coincides exactly with the optimal position within the region. However, this constraint considerably simplifies analysis, as selecting the best port reduces to a finite comparison problem. The principal analytical challenge lies in modeling the spatial correlation between ports in a tractable yet physically consistent manner, as no established correlation models readily enable closed-form analysis. In contrast, \glspl{CFAS} admits the possibility of positioning the \gls{FA} at the true optimal location within the region. The continuity of the spatial domain permits some closed-form analysis under well-established correlation models (e.g., Jakes model), without the discretization artifacts that complicate port-based analysis.

The positional reconfigurability central to \gls{FAS} operation can be achieved through several distinct hardware approaches, the most prominent of which are detailed below.
\begin{itemize}
    \item \textbf{Liquid metal} research initially inspired the \gls{FAS} concept. Conductive fluids, such as galinstan and eutectic gallium-indium, as well as ionic solutions, including sodium chloride and potassium chloride, have been proposed as liquid antennas \cite{wong_fluid_2021,psomas_continuous_2023}. These antennas can be repositioned within dielectric holders via microfluidic or electrowetting techniques.
    \item \textbf{Mechanically movable antennas} represent another implementation pathway for \glspl{FAS}. Although initially developed as a distinct research direction, such systems are now broadly subsumed under the FAS umbrella. They typically comprise a radiating element mounted on a flexible cable and repositioned continuously within a defined space by a micro-motor \cite{zhu_modeling_2024}.
    \item \textbf{Pixel-based switching} could also be used to realize a \gls{FAS}. Here, an upper layer of metal pixels is placed on an E-slot patch antenna and interconnected by \gls{RF} switches. Selectively activating these switches produces different current distributions, replicating the behavior of a physically repositioned antenna.
\end{itemize}
Due to the challenge of routing an \gls{RF} chain to a fluid radiating element, liquid metal antennas interface most naturally through fixed ports and thus suit \gls{DFAS} architectures. Similarly, pixel-based switching is inherently discrete, though dense arrays can approximate a  \gls{CFAS}. In contrast, mechanically movable antennas suit \glspl{CFAS}, as their direct connection to a \gls{RF} chain via flexible cable enables continuous repositioning.

\section{Recent Research Advances}
While the broader \gls{FAS} literature is extensive, the review here is confined to work directly pertinent to \glspl{CFAS}: analytical studies of \glspl{CFAS} themselves, comparisons between \glspl{CFAS} and \glspl{DFAS}, and spatially limiting studies of \glspl{DFAS}. Contributions are organized around four key themes.

\subsection{Outage Management}
The ability of \glspl{FAS} to reduce \gls{OP} is one of their principal advantages, and \glspl{CFAS} have been shown to be even more effective in this regard. The work in \cite{psomas_continuous_2023} investigated a \gls{SISO} system consisting of a \gls{FPA} transmitter communicating with a \gls{1D} \gls{CFAS}. Here, it was assumed that interfering channels dominate over any noise, and the \gls{LCR} and \gls{AFD} of the \gls{SIR} were derived. From these results, an approximation to the upper tail of the \gls{OP} was derived by decomposing it into two components: the probability that the spatial process exceeds the threshold at the initial position, and the probability of at least one threshold crossing occurring over the length of the \gls{FAS}, approximated by the product of the \gls{LCR} and the \gls{FAS} length. The derived \gls{OP} expression demonstrated that the \gls{CFAS} consistently outperforms the \gls{DFAS}, with the performance gap most pronounced at high thresholds where continuous spatial selectivity more effectively exploits the available diversity. These methods were extended in \cite{gayani} using a sojourn-time approximation for the entire distribution. More complex \gls{CFAS} architectures, including interference, movable arrays and mixed \gls{DFAS}/\gls{CFAS} layouts were handled using this approach.

The authors of \cite{new_fluid_2024} considered a \gls{DFAS} under the same scenario, examining the number of ports required to obtain performance at a level where the addition of further ports improved the \gls{OP} by less than a specified threshold. It was shown that the diversity gain is well approximated by the minimum of the number of ports and the rank of the channel covariance matrix, implying that port count increases are most beneficial until this rank is reached.

\subsection{Error Management}
The abilities of \glspl{CFAS} to enable ultra-reliable next-generation communications have been highlighted in recent papers by considering the limiting effects of increasing the number of ports in DFAS systems. In \cite{zhu_on_2025} and \cite{zhu_fluid_2025}, an asymptotic expression for the \gls{SER} of a \gls{SISO} system, involving a \gls{FPA} transmitter and \gls{1D} \gls{DFAS} receiver, was derived. The \gls{SER} was shown to be inversely proportional to the determinant of the port covariance matrix. When the determinant approaches unity, corresponding to  mutually uncorrelated channels, spatial diversity is maximized and the \gls{SER} is minimized. Conversely, as the correlation increases,  the effective rank (the number of non-negligible singular values) of the covariance matrix reduces and the system can no longer effectively resolve as many distinct spatial paths as there are ports. Additionally, \cite{zhu_fluid_2025} proved that as the number of ports tends to infinity, the effective rank  grows linearly with the normalized spatial bandwidth alone. Nevertheless, a higher port count improves spatial sampling, increasing the likelihood of selecting a port with a favorable channel realization.

The work in \cite{yang_ber_2026} considered a multi-\gls{FA} \gls{CFAS} at both the transmitter and receiver, under both \gls{ML} and \gls{ZF} detection. In both cases, an upper bound on the \gls{BER} was derived, shown to be proportional to the minimum Euclidean distance between transmit antennas, and the minimum singular value of the channel matrix. The former reflects the spatial separability of the transmit ports, while the latter captures the strength of the weakest spatial stream, with both measuring the effective available spatial \glspl{DoF}. Notably, the \gls{ZF} detector achieves performance equivalent to \gls{ML} when the channel matrix has a condition number of one, i.e., all singular values are equal and no spatial stream is disproportionately weak. This eliminates the noise enhancement penalty that typically disadvantages \gls{ZF}. The continuous positional freedom of \glspl{CFAS} allows the singular values to be finely adjusted toward a more balanced distribution, potentially enabling the implementation of the considerably simpler \gls{ZF} detector in scenarios where a \gls{DFAS} would preclude it.

\subsection{SNR and Capacity Improvement}
\Glspl{FAS} have also demonstrated the ability to improve \gls{SNR} and channel capacity. In \cite{wong_performance_2020}, an ergodic capacity bound was derived for a fundamental single-antenna \gls{1D} \gls{DFAS}. The ergodic capacity was found to saturate as the region size exceeded one wavelength, and the marginal capacity gain from adding further ports diminished as the port count increased.

In \cite{zhu_modeling_2024}, a field response model was developed for \glspl{CFAS}, considering a \gls{2D} single-antenna \gls{CFAS} at both the transmitter and receiver under deterministic and stochastic channel conditions. For deterministic channels, the maximum channel gain was found to vary periodically across the region, a behavior that enables the prediction of optimal \gls{CFAS} positions. For stochastic channels, as the region area tends to infinity, the capacity gain of the \gls{CFAS} over a \gls{FPA} grows proportionally with the number of channel paths. A higher number of channel paths leads to richer spatial variations, which can in turn be exploited by the \gls{CFAS}.

The work in \cite{new_channel_2025} investigated the impact of imperfect channel estimation on the average rate of a system employing a \gls{FPA} transmitter and a \gls{2D} \gls{CFAS} receiver. An electromagnetically compliant model was considered. It was shown that accurate channel reconstruction via maximum likelihood estimation requires sampling more frequently than the conventional Nyquist rate of half a wavelength, as the minimization of spectral leakage from the main lobe is necessary. To ensure this, a sampling interval of half a wavelength reduced by a term inversely proportional to the region size in wavelengths is required. Notably, even with imperfect channel reconstruction, the \gls{CFAS} was found to outperform \glspl{FPA}.

\subsection{Applications}
As a \gls{CFAS} can manage outages, reduce errors and improve the system \gls{SNR}, they can also be used to enable or enhance other next-generation wireless technologies. In \cite{gao_movable_2026}, \gls{2D} \glspl{CFAS} at the transmitter and receiver were employed for simultaneous wireless power transfer in the downlink and information transfer in the uplink. Joint optimization of \gls{FA} positions, time allocation, and power allocation yielded a system that outperformed a comparable \gls{FPA} system by almost 400\%. In \cite{chen_movable_2026}, a multi-antenna \gls{CFAS} was employed to enable \gls{ISAC}, with \gls{FA} positions and transmit beamforming jointly optimized to maximize the weighted sum of communication and sensing mutual information. It was found that \gls{CFAS}-enabled \gls{ISAC} required only half the number of antennas to match the performance of an \gls{FPA} array. Beyond the works discussed, \glspl{CFAS} have been integrated with a range of other next-generation technologies, including \glspl{UAV}, \glspl{RIS}, and \gls{CF} systems, underscoring their versatility as an enabling technology for future wireless networks.

\section{Dimensional Scaling Laws}
\begin{figure}
    \centering
    \includegraphics[width=\linewidth]{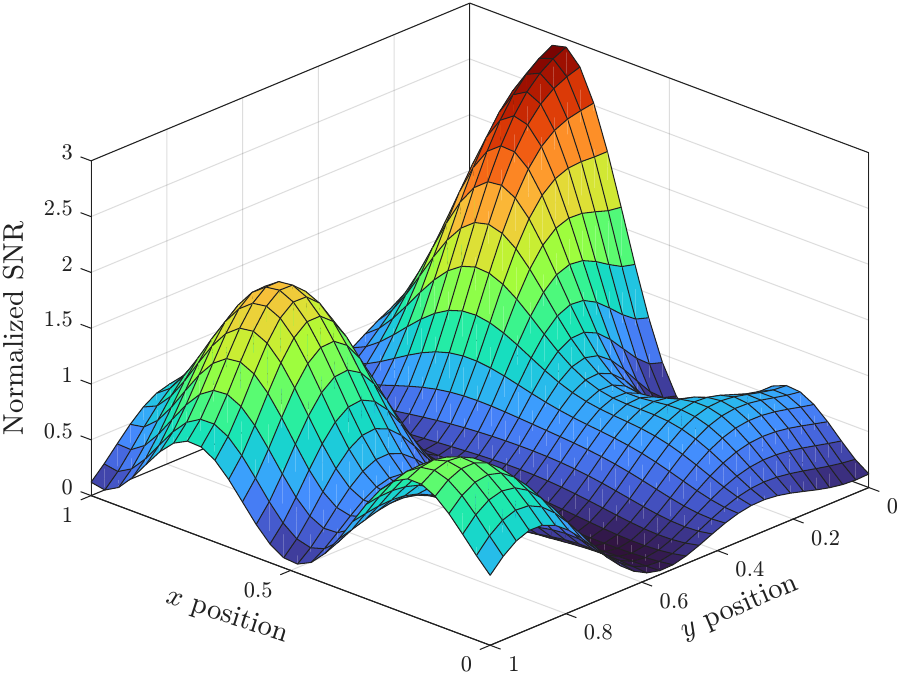}
    \caption{Example of the normalized \gls{SNR} experienced across a \gls{2D} square \gls{CFAS} with dimensions of one wavelength by one wavelength.}
    \label{fig:chan_resp}
\end{figure}

The characterization of \glspl{FAS} through closed-form expressions is notoriously difficult, with the upper tail of the \gls{SNR} distribution representing  one of the few analytically tractable metrics. A realization of the \gls{SNR} process for a single-antenna \gls{2D} \gls{CFAS} under the Jakes-correlated Rayleigh fading is shown in Fig. \ref{fig:chan_resp}. The supremum of this smoothly-varying process is attained by optimally positioning the \gls{FA} within the \gls{FAS} region, while the probability that the supremum exceeds some high \gls{SNR} threshold is referred to as the \gls{HSP}. Crucially, under correlated Rayleigh fading, the \gls{SNR} process of a single-antenna \gls{CFAS} follows a chi-squared distribution with two \glspl{DoF}, and it is this structure, combined with the focus on the upper tail, that makes closed-form analysis of the \gls{HSP} attainable. The \gls{HSP} thereby provides direct insight into the fundamental performance limits of \glspl{FAS} and the full extent of the spatial gains their deployment can offer.

As detailed in the preceding survey, \cite{psomas_continuous_2023} derived an asymptotically-exact expression for the upper tail of the supremum of the \gls{SIR} for a 1D \gls{CFAS} by employing the \gls{LCR} as an analytical tool. Here, the probability of the exceedance of a high threshold somewhere across the linear \gls{CFAS} region was expressed as the probability that the \gls{SIR} is initially above the threshold plus the expected number of upcrossings across the region. This same approach has been applied to obtain an asymptotically-exact expression for the \gls{HSP} of a single-\gls{UE} \gls{1D}  \gls{CFAS} in \cite{gayani}.

\begin{figure}
    \centering
    \includegraphics[width=\linewidth]{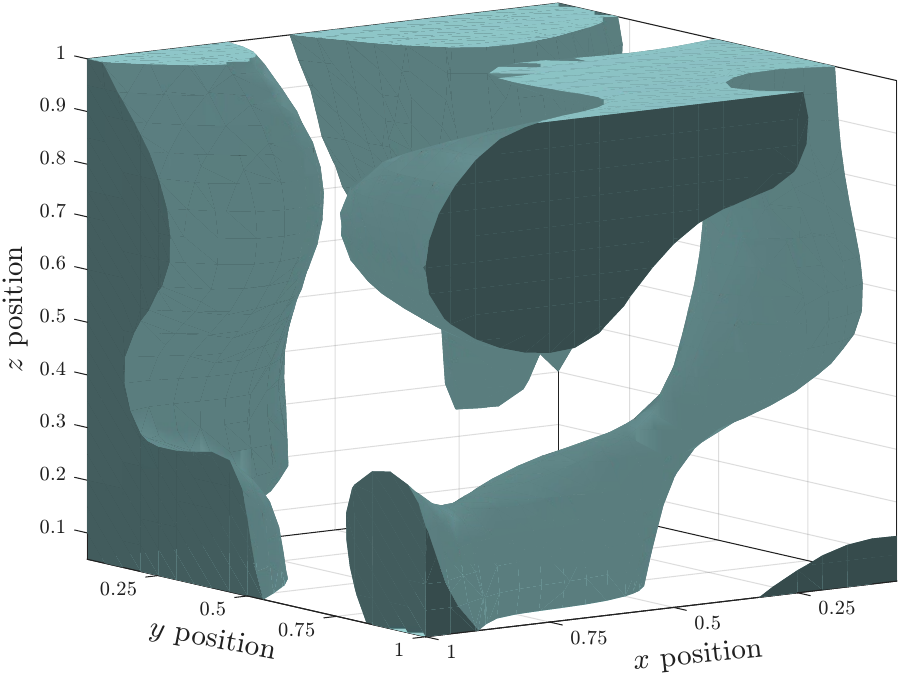}
    \caption{Example of the exceedance regions of a \gls{3D} cubic \gls{CFAS}.}
    \label{fig:3D_excur}
\end{figure}

This approach is well-suited to \gls{1D} systems, where exceedances occur over isolated intervals and the \gls{LCR} naturally quantifies their frequency. However, the extension to higher-dimensional \glspl{CFAS} is non-trivial. In \gls{2D}, the exceedance set comprises a scattered collection of irregular areas, and in \gls{3D}, an arrangement of disjoint volumes, as illustrated for a single cubic wavelength \gls{3D} \gls{CFAS} in Fig. \ref{fig:3D_excur}. In either case, the concept of a level crossing becomes geometrically ill-defined, precluding a straightforward extension of the \gls{LCR} approach.

Instead, the \gls{EEC} can be employed directly as a highly accurate approximation to the \gls{HSP}, as in \cite{smith_dimensional_2025}. The \gls{EEC} is a topological approach that counts exceedance regions, hence it functions analogously to the \gls{LCR} approach but is generalizable to any number of dimensions. This is crucial, as \gls{2D} and \gls{3D} \glspl{CFAS} require large numbers of simulated grid points to accurately characterize the region, making simulation-based approaches prohibitive for larger regions and necessitating closed-form expressions. The number of terms comprising the \gls{EEC} is equal to one more than the number of dimensions. Therefore, the contribution of each additional dimension is evident, and a dimensional scaling law can be obtained.

In \cite{smith_dimensional_2025}, by employing the \gls{EEC}, it was shown that the $n$-th dimension scales the \gls{HSP} of a single-\gls{UE} \gls{CFAS} with a hypercuboidal region under Rayleigh fading, such that
\begin{equation}
    P_{hs}^{(n)}\approx P_{hs}^{(n-1)}\left(1+T_n\sqrt{\frac{\lambda_2u_0}{2\pi}}\right), \label{eq:dimscale}
\end{equation}
where $P_{hs}^{(n)}$ is the \gls{HSP} of an $n$-dimensional \gls{CFAS}, $T_n$ is the length of the $n$-th dimension of the region, $\lambda_2$ is a constant that represents the variance of the channel derivative and is determined by the correlation model considered, and $u_0$ denotes the considered \gls{SNR} threshold. The scaling law admits a clean multiplicative form in which each newly added dimension results in a scaling term proportional to its length and to the square root of the \gls{SNR} threshold, rendering the joint dependence on geometry and exceedance level transparent.

The work in \cite{smith_dimensional_2025} considered solely a single-\gls{UE}, single-antenna uplink \gls{CFAS}. Here, we extend this by presenting a comparison with an uplink \gls{CFAS} involving two antennas  and serving two \glspl{UE} employing \gls{MMSE} combining at the receiver. While closed-form expressions are believed to be intractable in this multi-\gls{UE} setting, the scenario is important for understanding whether the dimensional scaling properties identified in the single-\gls{UE} case apply in a multi-\gls{UE} scenario. The high \gls{SINR} probability, defined as the probability that the \gls{SINR} of an arbitrarily selected \gls{UE} exceeds a high threshold, is contrasted with the \gls{SNR} of the single-\gls{UE} system across a range of dimensions and region sizes to ensure a fair comparison.

\subsection{Number of Dimensions}
\begin{figure}
    \centering
    \includegraphics[width=\linewidth]{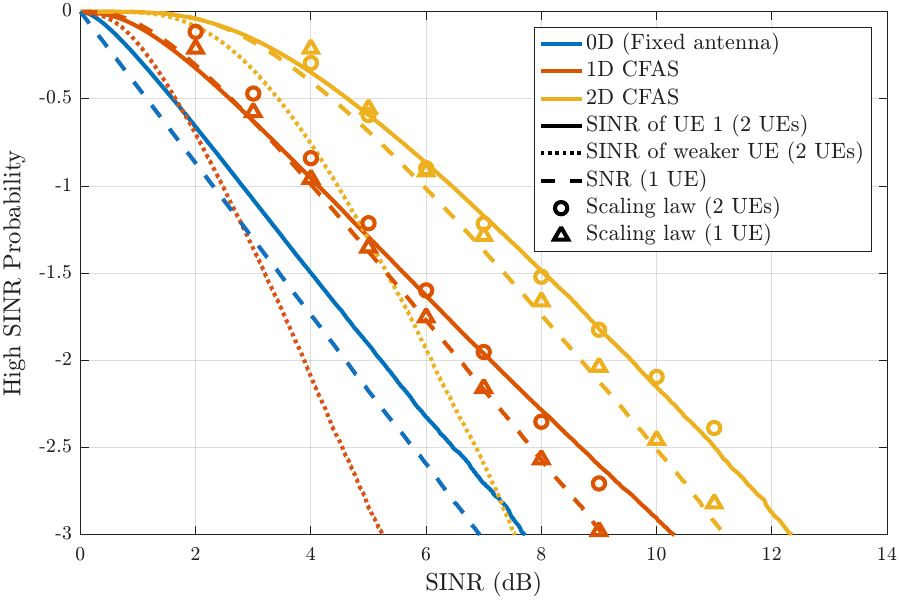}
    \caption{Comparison of \gls{HSP} performance for different \gls{CFAS} configurations.}
    \label{fig:HSP_dims}
\end{figure}

Figure \ref{fig:HSP_dims} compares the \gls{HSP} of a single-antenna, single-\gls{UE} system with high \gls{SINR} probabilities for a two-antenna, two-\gls{UE} system for 0D (fixed antenna), \gls{1D} and \gls{2D} \glspl{CFAS}. For 0D systems, it is assumed that antennas are spaced a wavelength apart and centered at the origin. The \gls{CFAS} region is a one wavelength line in \gls{1D} and a square with side lengths of one wavelength in \gls{2D}. For the single \gls{UE} case, the position of the \gls{FA} is selected as that which maximizes the \gls{SNR}. For the two-\gls{UE} case, the two \gls{FA} positions are selected to maximize the sum rate, rather than the \gls{SINR} of any individual user. The high \gls{SINR} probability of \gls{UE} 1 is then evaluated at this position, while the worst-case high \gls{SINR} probability is obtained by taking the lower of the two \glspl{SINR}. This reflects the performance of the weaker \gls{UE} under sum-rate-optimal \gls{FA} placement.  The dimensional scaling law in \eqref{eq:dimscale} has been applied to the fixed-antenna results for both the HSP and the high SINR probability of UE 1 to give approximate results in 1 and \gls{2D}. While the scaling law yields elegant closed-form results in the single-\gls{UE} setting, whether similar behavior occurs in the multi-\gls{UE} setting remains unknown. As closed-form analysis is believed to be intractable in this case, the single-\gls{UE} expression is instead applied to a two-antenna, two-\gls{UE} system as an initial study, with the aim of motivating further research into whether similar scaling behavior could emerge more generally.

Figure \ref{fig:HSP_dims} demonstrates that the scaling law in \eqref{eq:dimscale} remains accurate for a two-\gls{UE}, two-antenna \gls{CFAS}. This indicates that, for this system, the influence of additional dimensions on the high \gls{SINR} probability is dominated by the \gls{CFAS} size and  \gls{SINR} threshold, as is the case for a single-\gls{UE} \gls{CFAS}. The high \gls{SINR} probability of an arbitrarily selected \gls{UE} in the two-\gls{UE} scenario exceeds that of the single-\gls{UE} case, indicating that the gains afforded by the additional \gls{FA} outweigh the losses due to inter-user interference. This is particularly notable given that sum-rate maximization may, in some instances, yield a weak \gls{SINR} for an individual \gls{UE}, yet the per-user high \gls{SINR} probability remains superior. The gap between the \gls{HSP} and the high \gls{SINR} probability of \gls{UE} 1 at very high thresholds widens with increasing dimensionality. In the single-\gls{UE} case, a single position must be found at which the \gls{SNR} exceeds the threshold, whereas the additional \gls{FA} in the two-\gls{UE} case introduces an extra degree of freedom, enabling the contributions of both antennas to combine and jointly exceed the threshold. Additionally, the high \gls{SINR} probability of the weaker \gls{UE} in \gls{1D} is outperformed by the \gls{FPA} in both the lower and upper tails. Since the minimum \gls{SINR} across users is considered, sum-rate maximization may prioritize the stronger \gls{UE}, at times leaving the weaker \gls{UE} with an \gls{SINR} below that of a single-\gls{UE} served by a \gls{FPA} at the origin. This effect is further compounded in the upper tail, where it is inherently difficult for the minimum \gls{SINR} to exceed a very high threshold. Consequently, there exist instances in which a \gls{FPA} outperforms a \gls{CFAS} for a non-prioritized \gls{UE}. Nevertheless, the spatial diversity afforded by the \gls{CFAS} yields far superior overall performance than a \gls{FPA}, even with the presence of inter-user interference.

\subsection{CFAS Region Size}
\begin{figure}
    \centering
    \includegraphics[width=\linewidth]{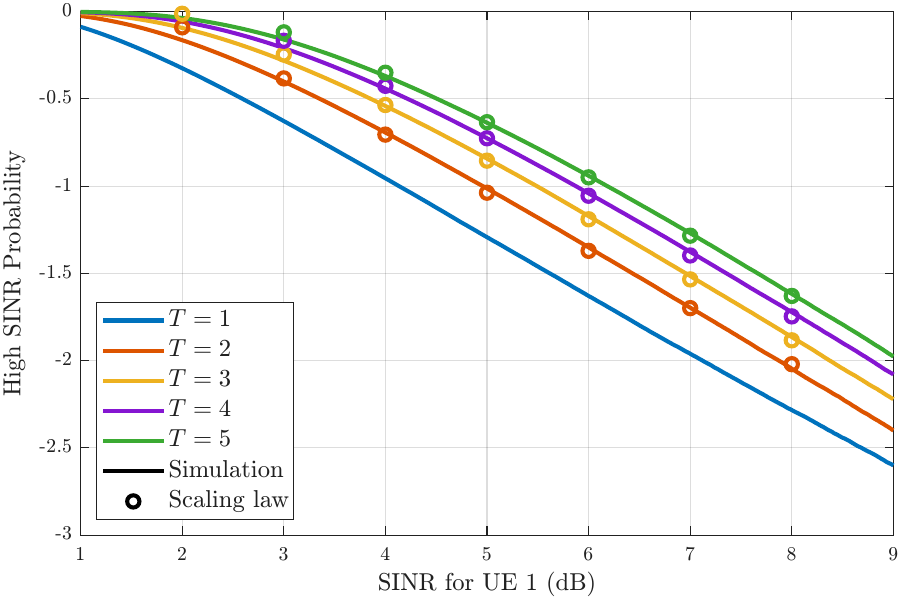}
    \caption{High \gls{SINR} probabilities of \gls{UE} 1 for \gls{1D} two-antenna \glspl{CFAS} of varying lengths.}
    \label{fig:HSP_T2}
\end{figure}

Figure \ref{fig:HSP_T2} examines the effect of size on the high \gls{SINR} probability. As an example, we consider a \gls{1D} \gls{CFAS}, where two antennas serve two \glspl{UE}. As before, antenna positions are set to maximize the sum rate, and the high \gls{SINR} probability of \gls{UE} 1 is shown for \gls{CFAS} lengths ranging from one to five wavelengths. The validity of the scaling law in \eqref{eq:dimscale} is again assessed, where results for a given \gls{CFAS} length are obtained by scaling those of the previous length by the ratio of \eqref{eq:dimscale} evaluated at the new length to \eqref{eq:dimscale} evaluated at the previous length.

Figure \ref{fig:HSP_T2} shows that the high \gls{SINR} probability increases with \gls{CFAS} length, reflecting the additional spatial diversity afforded by a larger region and the consequent increase in the likelihood that a position exists where the \gls{SINR} exceeds the threshold. These gains diminish with increasing length, as the best available position is already likely to be highly favorable, and the marginal benefit of further extending the region decreases accordingly. The scaling law in \eqref{eq:dimscale} is seen to be highly accurate across all lengths considered. As only the \gls{CFAS} length varies between curves, the ratio of consecutive evaluations of \eqref{eq:dimscale} depends solely on the two lengths considered. This ratio is monotonically decreasing with length, so the scaling law directly captures the diminishing gains in high \gls{SINR} probability observed as the \gls{CFAS} length increases.

\section{Open Research Problems}
As \glspl{FAS}, and particularly \glspl{CFAS}, remain an emerging technology, several significant research challenges persist. The most pertinent of these are discussed below.

\subsection{Analysis of the Outage Probability Lower Tail}
While highly accurate approximations to the \gls{OP} of \glspl{CFAS} have been obtained, such as in \cite{psomas_continuous_2023} and \cite{smith_dimensional_2025}, these have focused exclusively on the upper tail of the distribution. The lower tail is arguably the more critical region, as it characterizes the ability of the \gls{CFAS} to recover weak channel realizations and is therefore directly relevant to reliability and outage performance in adverse conditions. However, the lower tail is significantly more difficult to characterize analytically, as it corresponds to the probability that the supremum of the \gls{SNR} over the full range of possible antenna positions remains below a specified threshold. These quantities are known as \textit{small ball probabilities}, and are difficult to obtain in closed form. This is especially pertinent for \gls{2D} and \gls{3D} \glspl{CFAS}, where the density of sample points required to accurately simulate the supremum grows rapidly with dimensionality, rendering purely numerical approaches computationally prohibitive and making closed-form analytical results crucial. Note that the OP approximations in \cite{gayani} are useful for the main body of the distribution, but have limited accuracy in the lower tail.

\subsection{Impacts of Latency}
As discussed earlier in this article, mechanically movable antennas are a key candidate for the implementation of \glspl{CFAS}, due to their ability to be directly connected to a flexible \gls{RF} chain. However, this will require the antenna to physically move to an updated position, which will incur a latency period. Accurate characterization of this delay is essential both for informing position selection algorithms and for enabling fair performance comparisons with near-instantaneous implementation methods, such as pixel-based switching.

\subsection{Channel State Information Acquisition Methods}
The work in \cite{new_channel_2025} provided crucial insights into the required sampling interval for accurate channel reconstruction in \glspl{CFAS}. However, open questions remain regarding how \gls{CSI} will be physically acquired and how frequently it must be updated. As discussed above, the delays incurred while shifting a mechanically movable antenna will provide a number of challenges, one of which will be its impact on channel estimation. In rapidly varying channels, measurements taken at early positions may become outdated before the sweep across the full range is complete, undermining the validity of the estimated channel response. This raises the important question of what environments \glspl{CFAS} will be viable in: specifically, whether they will operate effectively in rapidly varying channels, or whether they are better suited to highly correlated or semi-static propagation environments.

\section{Conclusions}

This article examined the spatial performance limits of \glspl{FAS}, with particular emphasis on the \gls{HSP} as a tractable and informative metric. A survey of recent contributions across outage management, error performance, \gls{SNR} improvement, and emerging applications was conducted. Next, the dimensional scaling law for the \gls{HSP} was investigated beyond its original single-UE setting. It was demonstrated that the scaling law retains its accuracy for a \gls{CFAS} involving multiple \glspl{UE} and antennas, indicating that the dependence of the high \gls{SINR} probability on region size and threshold is robust to the introduction of multiple users and inter-user interference. Furthermore, the spatial diversity afforded by \glspl{CFAS} was shown to yield superior performance relative to  fixed-position antennas. These findings represent an important step toward the analytical characterization of more complex \gls{CFAS} configurations. While significant advances in the area of \glspl{CFAS} have been made, many open problems remain, including the analytical characterization of the \gls{OP} lower tail, the rigorous modeling of positioning latency in mechanically movable implementations, and the development of robust \gls{CSI} acquisition strategies suited to the continuous spatial domain. Addressing these challenges will be essential to realizing the full potential of \glspl{CFAS} in next-generation wireless networks.

\bibliographystyle{IEEEtran}
\bibliography{references}

\begin{IEEEbiographynophoto}{Amy S. Inwood}
received her Ph.D. degree in electrical and electronic engineering from Te Whare Wānanga o Waitaha \textbar{} University of Canterbury, NZ, in 2024. She is now a Research Fellow at the Centre for Wireless Innovation, Queen's University Belfast, U.K.
\end{IEEEbiographynophoto}

\begin{IEEEbiographynophoto}{Peter J. Smith}
    received the Ph.D degree in Statistics from the University of London, U.K., in 1988. In 2015 he joined Victoria University of Wellington as Professor of Statistics. He is an Honorary Professor in the School of Electronics, Electrical Engineering and Computer Science, Queen's University Belfast.
\end{IEEEbiographynophoto}

\begin{IEEEbiographynophoto}{Rajitha Senanayake}
    received her Ph.D. degree in electrical and electronics engineering from the University of Melbourne, Australia, in 2015. From 2015 to 2016, she was with the Department of Electrical and Computer Systems Engineering, Monash University, Australia. Currently, she is a Senior Lecturer with the Department of Electrical and Electronics Engineering, University of Melbourne. She was a recipient of the Australian Research Council Discovery Early Career Researcher Award.
\end{IEEEbiographynophoto}

\begin{IEEEbiographynophoto}{Michail Matthaiou}
    obtained his Ph.D. degree from the University of Edinburgh, U.K. in 2008. He is currently a Professor of Communications Engineering and Signal Processing and Deputy Director of the Centre for Wireless Innovation (CWI) at Queen’s University Belfast, U.K. He is also an Eminent Scholar at the Kyung Hee University, Republic of Korea. He currently holds the ERC Consolidator Grant BEATRICE (2021-2027) focused on the interface between information and electromagnetic theories. He has received the prestigious 2023 Argo Network Innovation Award, the 2019 EURASIP Early Career Award and the 2018/2019 Royal Academy of Engineering/The Leverhulme Trust Senior Research Fellowship.
\end{IEEEbiographynophoto}

\end{document}